\documentclass[aps,twocolumn,float,prl,psfig]{revtex4}

\usepackage{color}

\newcommand{\add}[1]{\textcolor{red}{#1}}

\input epsf
\usepackage{graphicx}
\newcommand{\beq}{\begin{equation}}
\newcommand{\beqa}{\begin{eqnarray}}
		  \newcommand{\eeq}{\end{equation}}
\newcommand{\eeqa}{\end{eqnarray}}

\newcommand{\gsim}{\gtrsim}
\newcommand{\psim}{\mbox{\raisebox{-1.0ex}{$~\stackrel{\textstyle \propto}
{\textstyle \sim}~$ }}}
\newcommand{\vect}[1]{\mbox{\boldmath${#1}$}}
\newcommand{\lmk}{\left(}
\newcommand{\rmk}{\right)}

\newcommand{\lkk}{\left[}
\newcommand{\rkk}{\right]}
\newcommand{\lla}{\left\langle}

\newcommand{\rra}{\right\rangle}

\newcommand{\vex}{{\vect x}}
\newcommand{\veh}{{\vect h}}
\newcommand{\vue}{\hat{\vect e}}

\newcommand{\ven}{\vect n}

\newcommand{\vep}{{\vect p}}

\newcommand{\veb}{{\bf e}}

\begin{document}
\title{Measuring Parity Asymmetry  of Gravitational Wave Backgrounds with a Heliocentric Detector Network  in the  mHz Band
 } 
%
%
%
\author{Naoki Seto }
\affiliation{Department of Physics, Kyoto University, 
Kyoto 606-8502, Japan
}
\date{\today}
%
%
%
%
%
%
\begin{abstract}
 
We discuss exploration for isotropic gravitational wave backgrounds around 1\,mHz by correlation analysis, targeting both  parity odd and even polarization modes. Even though the space interferometer LISA alone cannot probe the two modes due to  cancellations,  the outlook is being changed drastically by the strong development of other space detectors such as Taiji.   In fact, a  heliocentric interferometer network holds a  preferable geometrical symmetry {illuminated by a virtual sphere off-center from the Sun}. By  utilizing an internal symmetry of data streams, we  can optimally decompose the odd and even parity modes at  correlation analysis.    By simultaneously using LISA and Taiji for 10 years, our sensitivity to the two modes  could reach $\sim 10^{-12}$ in terms of the normalized energy density. 

\end{abstract}

\maketitle

\section{introduction}

Given the high penetration power of gravitational waves, a stochastic gravitational wave background could be a very important fossil from the early universe 
for studying an extreme physical state \cite{Romano:2016dpx,Caprini:2018mtu}.   For cosmology, we would like to primarily search for the monopole components of a background, since our observed universe is nearly isotropic at large angular scale.

The Stokes $V$ parameter characterizes the asymmetry between the amplitudes of the right- and left handed polarized  waves, and is proportional to $\Pi\Omega_{\rm GW}$ in the case of gravitational waves.  Here,  $\Omega_{\rm GW}$  is the normalized energy  density and $\Pi$ is the polarization degree ($\Pi=\pm1$ for 100\% right-/left-handed waves).  By measuring the $V$ parameter, we can probe a parity violation process.  Indeed, there are many cosmological arguments on the circularly polarized  gravitational waves, including their potential roles for leptogenesis  \cite{Alexander:2004us}    and  their production coupled with Chern-Simon terms during inflation (see e.g. \cite{Adshead:2012kp} for chromo-natural inflation, \cite{Obata:2014loa,Maleknejad:2012fw} for its extensions  and \cite{Takahashi:2009wc} for a quantum gravity effect at a Lifshitz point) and phase transitions \cite{Kahniashvili:2005qi,Ellis:2020uid}.  {For example, depending on model parameters, due to a gauge field, a nearly 100\% polarized ($|\Pi|\sim 1$) background might be generated above $10^{-11}$\,Hz with the amplitude $\Omega_{GW}\gsim 10^{-12}$ \cite{ Obata:2014loa}. }  Therefore a detection of non-vanishing $V$ parameter will  have significant impacts on fundamental physics. 

By correlating data streams of  noise independent-interferometers, we can directly measure the Stokes $V$ parameter \cite{Seto:2007tn,Seto:2008sr,Smith:2016jqs}  (see also \cite{Lue:1998mq,Contaldi:2008yz,Inomata:2018rin} for CMB analysis). For example,  in the 10-1000Hz band,  using the current generation ground-based network for a few years, we will be able to detect the $V$ parameter corresponding to  $|\Pi|\Omega_{\rm GW} \sim 10^{-8}$ \cite{Seto:2008sr}.

The band around 0.1m-1\,Hz will be explored by space interferometers. The LISA project has a history of over 20 years \cite{lisa0,lisa}, and its pathfinder mission recently made an  impressive success \cite{Armano:2018kix}.  From its triangle  constellation,  we can generate multiple noise-independent data channels \cite{Prince:2002hp}.  But, unfortunately, LISA is totally insensitive to the isotropic component of the  $V$ parameter, due to the exact cancellation resulting from  the mirror symmetry at the interferometric  plane \cite{Seto:2006hf}. With LISA alone, we can merely observe anisotropic pattern (e.g. $l=1$ and 3 harmonics) of the $V$ parameter \cite{Seto:2006hf,Domcke:2019zls} ({see also} \cite{Belgacem:2020nda}).  {In fact, we also have an independent cancellation mechanism related to  the symmetry of the data channels, and even the energy density $\Omega_{\rm GW}$ cannot be measured by correlating LISA\rq{}s data streams ({see} also \cite{Hogan:2001jn,Tinto01,Adams:2010vc,Romano:2016dpx} for estimating $\Omega_{\rm GW}$ with the Sagnac modes).  The future space plans  such as BBO and DECIGO are designed to use multiple triangles  for correlation analysis. By relatively tilting orbits of two triangles, we can measure the even and odd parity modes down to $\Omega_{\rm GW}\sim \Omega_{\rm GW}|\Pi|\sim 10^{-16}$ \cite{Seto:2006dz}, but these missions will be available  much later than LISA.  }

However, nowadays, two other projects (TianQin \cite{Luo:2015ght} and Taiji \cite{taiji,Ruan:2019tje}) are actively propelled  in the mHz band,  both aiming operation around 2035, similar to LISA (see Fig. 1). Therefore, in the mHz band,  we now have an increased chance to study  an isotropic background by correlating LISA and Taiji/TainQin, without  being hampered by the various signal cancellations. This paper is the first quantitative study on this issue.

Since  observation of a gravitational wave background  is intrinsically geometrical measurement, it is crucially important to see through the underlying symmetry of the network.  From this standpoint, we limit our analysis to  a network composed by heliocentric interferometers (more specifically LISA and Taiji), paying special attention to the measurement of the  two parity modes.  In fact, a heliocentric  network could have  favorable  geometrical symmetries that will allow us to easily make an optimal parity decomposition of a background. 

\begin{figure}
 \includegraphics[width=8.cm]{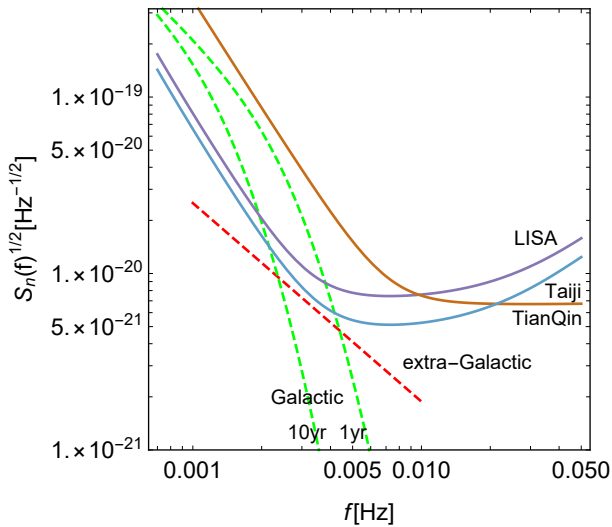}
 \caption{Solid curves: the noise spectra of proposed space interferometers  (LISA, Taiji and TianQin) for single data channels ($A$ and $E$ types).  The green dashed  curves are  estimation for the Galactic foreground (based on \cite{Cornish:2018dyw}) with observational time $T_{\rm obs}=1$ and 10\,yr. 
The red dashed curve shows the pessimistic model for extra-Galactic white dwarf  confusion noise in \cite{Farmer:2003pa} with $\Omega_{\rm GW}(f)\simeq 0.95 \times 10^{-12} (f/{\rm 1\,mHz})^{3/4}$  at 1\,mHz-10\,mHz.
} \label{fig:1}
\end{figure}


\section{Symmetries of the system}
Here, for a heliocentric detector network,  we discuss two symmetries that are  important for the parity decomposition. To the author's best knowledge,  the first one had been never covered in the literature. The second one has been known (see e.g. \cite{Seto:2004ji}), but can  play a particularly interesting role when coupled with the first one.
\subsection{global symmetry}

LISA has a heliocentric orbit moving $20^\circ$  behind the Earth. Its three spacecrafts nearly keep a regular triangle with the arm length $L=2.5\times 10^6$\,km \cite{lisa}. This can be achieved by initially adding small eccentricities and inclinations to the spacecrafts, and the detector plane is  resultantly  inclined to the  ecliptic plane by $60^\circ$. In Fig. 2,  with the gray belt, we illustrate the envelope of the detector plane. The middle of the belt is on the ecliptic plane and corresponds to the orbital line of the barycenter of each triangle,  with the radius $R_E=1.5\times 10^8{\rm \,km}=1.0$\,AU.  {The triangle is also spinning on the belt with a period of 1\,yr (the so called cartwheel rotation).}

Taiji moves $20^\circ$ ahead of the Earth with its arm length $L=3.0\times 10^6$\,km \cite{taiji} (see also \cite{Cornish:2001bb}), and shares the envelope with LISA (not opposite direction). The separation between LISA and Taiji is $D=2R_E\sin(20^\circ)=1.0\times 10^8$\,km.  TianQin has a geocentric orbit, and should be analyzed separately.

Here, it should be noticed that the gray belt in Fig. 2 contacts with a  sphere (hereafter \lq\lq{contact sphere}\rq\rq{}) of  radius $R_{\rm C}=(\sqrt3/2)^{-1}R_E=1.15$\,AU.   Interestingly, the contact sphere is not \lq\lq{}heliocentric\rq\rq{}, but its center $\rm O_C$ is at 0.58\,AU north of the ecliptic plane.  As a sphere is highly symmetric object and search for gravitational wave backgrounds is a geometrical measurement, it would be advantageous to view the LISA-Taiji system in relation to the contact sphere.  
For example, from its center $\rm O_C$, LISA and Taiji are separated by 
\beq
\beta=2\sin^{-1}[D/(2R_{\rm C})]=34.5^\circ.
\eeq

In order to quantify the relative orientation of LISA and Taiji interferometers  on their detector planes (as discussed in the next subsection), we introduce a curve connecting the positions of the two observatories.  Considering the existing symmetry of the system, the primary choice will be  the great circle (geodesic) on the contact sphere, not the orbital line on the ecliptic plane.

\begin{figure}
 \includegraphics[width=7.4cm,angle=270,clip]{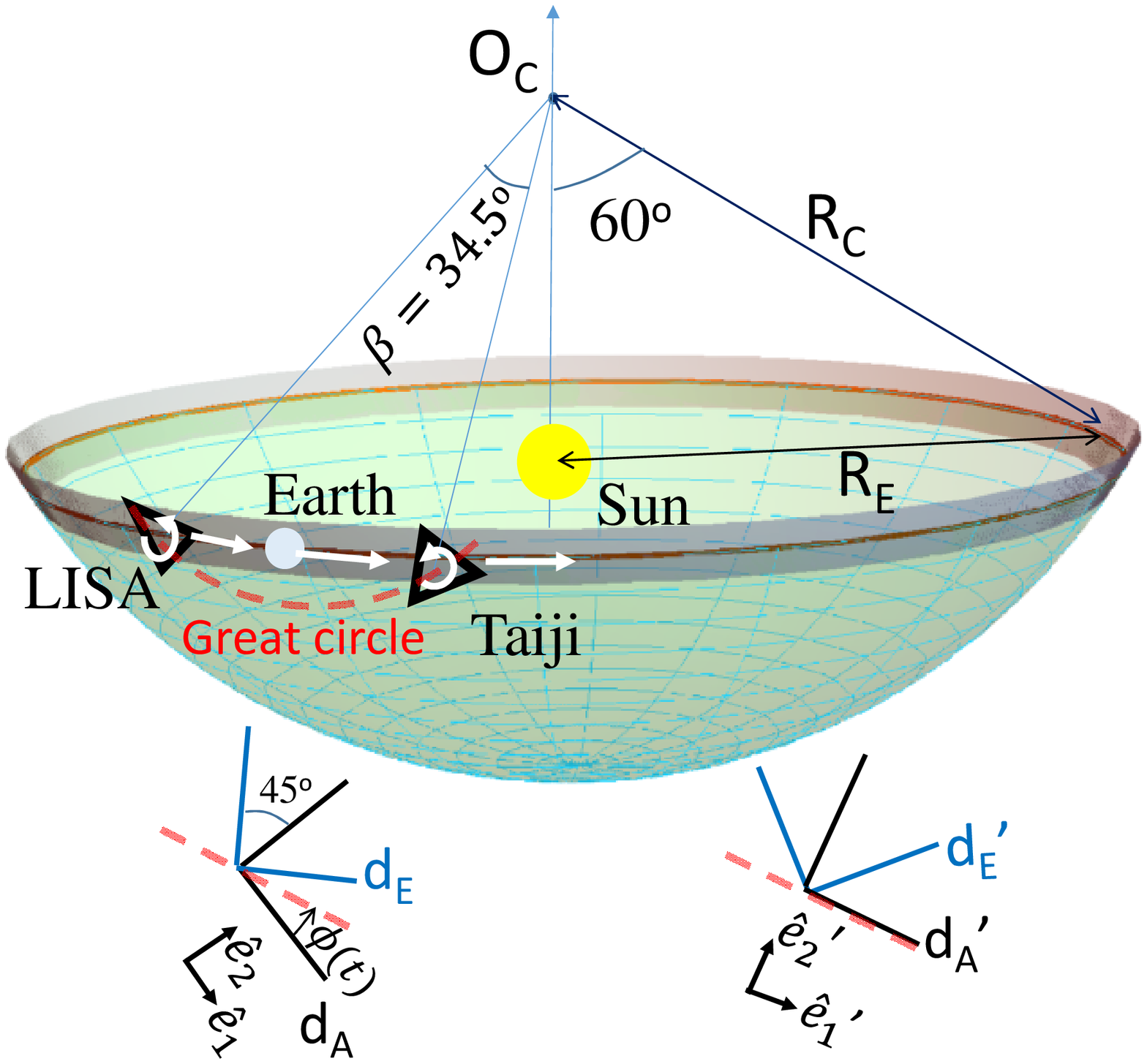}
 \caption{ (Upper panel) The gray belt shows the envelope of the detector planes of the two  heliocentric triangle interferometers, LISA and Taiji. The orange line is the orbital line of the barycenter of each triangle. The green surface is a part of the virtual  sphere of radius $R_{\rm C}=1.15$\,AU contacting the envelope belt. From  the center $\rm O_C$, the angular separation between LISA and Taiji is $\beta=34.5^\circ$ with the great circle shown with the dashed  curve. The triangles are spinning on the belt (cartwheel motion).  
(Lower left panel)  The orientation of the detector tensors ${\bf d}_A$ and ${\bf d}_E$ that is attached to the LISA\rq{}s triangle and  characterized by the co-spinning  orthonormal basis $(\vue_1,\vue_2)$.  We define  $\phi(t)$ for  the time dependent miss-alignment angle  relative to the great circle.
(Lower right panel) The orientation of the virtual detector tensors ${\bf d}_A\rq{}$ and  ${\bf d}_E\rq{}$ associated with the aligned basis  $(\vue_1\rq{},\vue_2\rq{})$. For the mirror transformation at the plane containing the great circle and $\rm O_C$, we have $\vue_1\rq{}\to \vue_1\rq{}$ and $\vue_2\rq{}\to -\vue_2\rq{}$ (accordingly ${\bf d}_A\rq{}\to {\bf d}_A\rq{}$ and ${\bf d}_E\rq{}\to- {\bf d}_E\rq{}$). }
 \label{fig:2}
\end{figure}


\subsection{internal symmetry}

Next we discuss symmetry within  the triangle of the LISA constellation (essentially the same for Taiji).   We can compose three interferometers at the three vertexes that are assumed to be equivalent (as  in the  standard literature)  \cite{Prince:2002hp}.  However, their data have correlated noises. Using the symmetry  of the vertexes, we can make three noise independent data channels ($A,E,T$) as linear combinations of the three original data \cite{Prince:2002hp}.   The {Sagnac-like} $T$ channel has a  negligible sensitivity to gravitational waves in the low frequency regime $f<c/(2\pi L)\sim 20$\,mHz (with $L\ll D$), and  can be used when  measuring the instrumental noise level for the spectral estimation of a gravitational wave  background  \cite{Hogan:2001jn,Tinto01,Adams:2010vc,Romano:2016dpx}.  For correlation analysis,  we thus consider to use the $A$ and $E$ channels below, applying the low frequency approximation.    Reflecting the original symmetry of the three vertexes, their  noise spectra are identical \cite{Prince:2002hp}, but, at the same time, using their correlation,  we cannot measure the monopole pattern irrespective of the parity modes (see e.g. \cite{Seto:2006hf}).

The $A$ and $E$ channels of LISA can be effectively  regarded as responses of two L-shaped  interferometers with orientation difference $45^\circ$, as shown in the lower left panel of Fig. 2 (see e.g. \cite{Romano:2016dpx}). They are attached to the LISA\rq{}s spinning triangle.  Here, to characterize interferometric responses,  we introduce the detector tensors ${\bf d}_A$ and ${\bf d}_E$. For the $A$ channel, we have ${\bf d}_A=(\vue_{1}\otimes\vue_{1}-\vue_{2}\otimes\vue_{2} )/2 $ with the unit co-spinning vectors $(\vue_{1},\vue_{2})$ for its two arm directions.  For the $E$ channel, using the same unit  vectors, we have ${\bf d}_E=(\vue_{1}\otimes\vue_{2}+\vue_{2}\otimes\vue_{1})/2$.   The combination $({\bf d}_A,{\bf d}_E)$ forms the orthogonal basis for the detector tensors on the instantaneous detector  plane. 
Note that the orientation of the detector tensors ${\bf d}_A$ and ${\bf d}_E$ are not aligned with the great circle (given the cartwheel spin rotation). We put the time-dependent  miss-alignment angle by $\phi(t)$ as shown in Fig. 2.

 Now we virtually rotate the basis  $(\vue_{1},\vue_{2})$ commonly by the angle $\phi(t)$ so that $\vue_1\rq{}$ is parallel to the great circle and respects the global symmetry of the network.  We call the corresponding virtual  detector tensors by ${\bf d}_{A\rq{}}=(\vue_{1}\rq{}\otimes\vue_{1}\rq{}-\vue_{2}\rq{}\otimes\vue_{2}\rq{} )/2$ and ${\bf d}_{E\rq{}}=(\vue_{1}\rq{}\otimes\vue_{2}\rq{}+\vue_{2}\rq{}\otimes\vue_{1}\rq{} )/2$ (see the lower right panel in Fig. 2).
 By tensorial calculations, we can directly confirm the relation (e.g. \cite{Seto:2004ji})

\beq
\left(
    \begin{array}{c}
      {{\bf d}_{A\rq{}}} \\ 
        {{\bf d}_{E\rq{}}}
    \end{array}
  \right)
= \left(
    \begin{array}{cc}
      \cos 2\phi(t)& \sin 2\phi(t) \\ 
       -\sin2\phi (t) & \cos2\phi(t)
    \end{array}
  \right) \left(
    \begin{array}{c}
      {\bf d}_{A} \\ 
        {\bf d}_{E}
    \end{array}
  \right) \label{ae}
\eeq
with the factor 2 reflecting the spin-2 nature. 
This means that,  by linearly combining the LISA\rq{}s original data channels $A$ and $E$ in the same manner as Eq. (\ref{ae}), we can   actually obtain the  data  channels $A\rq{}$ and $E\rq{}$ whose detector tensors are the virtual ones ${\bf d}_{A\rq{}}$ and ${\bf d}_{E\rq{}}$.
 The new data set $(A\rq{},E\rq{})$ have the same information content and noise spectrum as the original set $(A,E)$, still without correlation (as easily confirmed).  We can make a similar adjustment for Taiji.

 We hereafter call the (virtually generated) aligned  channels  by $(A_{\rm L},E_{\rm L})$ for LISA and $(A_{\rm T},E_{\rm T})$ for Taiji.   In the next section, we consider the four inter-detector combinations  $A_{\rm L}$-$A_{\rm T}$, $E_{\rm L}$-$E_{\rm T}$, $A_{\rm L}$-$E_{\rm T}$  and $E_{\rm L}$-$A_{\rm T}$.

 We should notice that, for the mirror transformation with respect to  the plane containing the great circle and $\rm O_C$, the unit vectors are transformed as $\vue_{1}\rq{}\to  \vue_{1}\rq{}$ and $\vue_{2}\rq{}\to- \vue_{2}\rq{}$ both for LISA and Taiji (see the lower right panel in Fig. 2).  Then the detector tensors  have  even parity for $(A_{\rm L},A_{\rm T})$  and  odd parity for $(E_{\rm L},E_{\rm T})$ (see the definitions of ${\bf d}_A\rq{}$ and ${\bf d}_E\rq{}$ above).    These properties would be essential for the optimal parity decomposition of a gravitational wave background.

Generally speaking, for two triangle detectors given at positions $\vep_{\rm i}$  (i=1,2) with associated normal vectors $\ven_{\rm i}$,  the above symmetric  mirror transformation can be  applicable, only if the three vectors $\vep_1-\vep_2$, $\ven_1$ and $\ven_2$ are linearly dependent.  In this sense, the gray envelope in Fig. 2 is geometrically special.

 \section{gravitational wave background}

\subsection{monopole pattern}
We use the Fourier decomposition of the metric perturbation induced by gravitational waves as
\beq
\veh(t,\vex)=\sum_{P=R,L} \int^{\infty}_{-\infty} df \int_{S^2} d\ven~
h_P(f\ven)  \veb^P_{}(\ven) e^{2\pi i f (\ven \cdot \vex-t) },
\eeq
where we adopted the right- and left-handed polarization bases $\veb^{R,L}(\ven)$ with the unit propagation vector $\ven$. They are given by the familiar linear polarization bases $\veb^{+,\times}$ as 
\beq
\veb^R=(\veb^++i \veb^\times)/\sqrt{2},~~~\veb^L=(\veb^+-i \veb^\times)/\sqrt{2} \label{+x}.
\eeq

Our target in this paper is a stationary and isotropic gravitational wave background. For the monopole components, we can generally write  \cite{Seto:2006hf,Seto:2008sr,Smith:2016jqs}
\beqa
\left(
  \begin{array}{c}
     \lla h_R(f\ven) h_R(f'\ven')^* \rra \\
     \lla h_L(f\ven) h_L(f'\ven')^* \rra 
  \end{array}
 \right)
=\frac{ \delta_{{\ven,\ven'}}\delta_{f,f'} }{8\pi}
\left(
  \begin{array}{c}
     I(f)+V(f) \\
     I(f)-V(f) 
  \end{array}
 \right) \label{rl}
\eeqa
with the Stokes parameters $(I,V)$ and the Delta functions. In Eq. (\ref{rl}), the parameter $I$ represents the total intensity of the background, while $V$ characterizes the asymmetry between the right- and left-handed waves, as mentioned earlier.   
Note that, for the monopole mode, we do not have the Stokes $Q$ and $U$ parameters related to linear polarization.

\subsection{correlation analysis}

In the Fourier space, reaction of an interferometer  $a$ (at $\vex=\vex_a$) to gravitational waves is given by
\beq
h_a(f)=\sum_{P=R,L} \int d\ven \,h_P(f\ven)({\bf d}_a:\veb^{P}) e^{2\pi i f \ven \cdot \vex_a }.
\eeq
{Here the colon $:$ represent the double contraction of the two tensors.}
Then correlation between two detectors $a$ and $b$ is given by $\lla h_a(f) h_b(f\rq{})^*\rra =C_{ab}(f)\delta_{f,f\rq{}}$.  Using Eqs. (\ref{+x}) and (\ref{rl}), we have
\beq
C_{ab}(f)=\frac{8\pi}5 \lkk \gamma_{Iab}(f)I(f)+\gamma_{Vab}(f) V(f) \rkk. 
\eeq
Here $\gamma_{Xab}$ are the overlap reduction functions  ($X=I,V$)
\beq
\gamma_{Xab}(f)=\frac5{8\pi} \int d\ven K_{Xab}(\ven) e^{2\pi i f \ven\cdot (\vex_a-\vex_b)}
\eeq
with the function $K_{Xab}(\ven)$ given by the beam patter functions ($F_a^{+,\times}(\ven)\equiv {\bf d}_a:\veb^{+,\times}(\ven)$) as  \cite{Seto:2006hf, Seto:2008sr}
\beq
K_{Iab}=F_a^+ F_b^++F_a^\times F_b^\times, ~K_{Vab}=-i(F_a^+ F_b^\times -F_a^\times F_b^+).
\eeq

\begin{figure}
 \includegraphics[width=8.5cm]{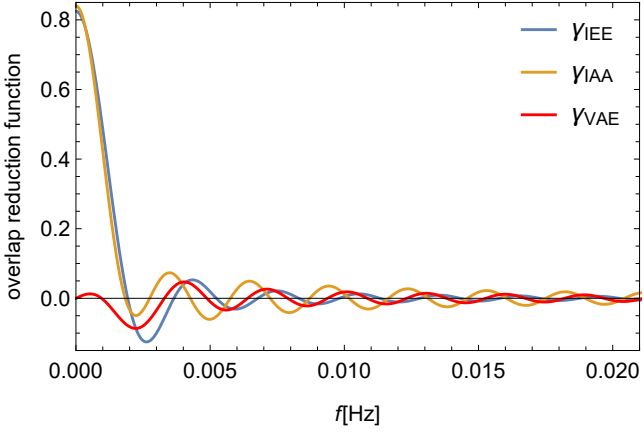}
 \caption{The overlap reduction functions for LISA-Taiji system.  We can make a complete separation between $I$ and $V$ modes by pairing the data channels $(A_{\rm L},E_{\rm L})$ and $(A_{\rm T},E_{\rm T})$. }
 \label{fig:3}
\end{figure}


The overlap reduction functions  describe the correlated responses of two interferometers to isotropic gravitational wave backgrounds. 
In \cite{Flanagan:1993ix,Allen:1997ad},  for ground based detectors, a simple analytic expression was derived  for $\gamma_{Iab}$, fully using the symmetry of the earth surface as a sphere (similarly $\gamma_{Vab}$ by \cite{Seto:2007tn}).  Quite remarkably, for the LISA-Taiji system, by virtually introducing the contact sphere ($\sim 10^{4.5}$ times larger than the Earth), we can directly apply the simple expressions for $\gamma_{Xab}$ originally provided for ground-based networks.

Moreover, the combinations made from the virtual data channels $(A_{\rm L},E_{\rm L})$ and  $(A_{\rm T},E_{\rm T})$ are parity eigenstates for the mirror transformation mentioned at the end of previous section.  In concrete terms, $A_{\rm L}$-$A_{\rm T}$  and $E_{\rm L}$-$E_{\rm T}$ have even parity, and $A_{\rm L}$-$E_{\rm T}$ and $E_{\rm L}$-$A_{\rm T}$ have odd parity, corresponding to the special types classified in \cite{Seto:2008sr}.   The even ones are sensitive only to the $I$ mode (blind to $V$), and the odd ones are opposite.  Accordingly, we can make a perfect decompositions of the $I$ and $V$ modes, using the present pairs.  For the overlap reduction functions, we can express\\
$(\gamma_I(f),\gamma_V(f))=(\Xi_1,0)$ for $A_{\rm L}$-$A_{\rm T}$,\\
$(\gamma_I(f),\gamma_V(f))=(\Xi_2,0)$ for $E_{\rm L}$-$E_{\rm T}$,\\
$(\gamma_I(f),\gamma_V(f))=(0,\Xi_3)$ for $A_{\rm L}$-$E_{\rm T}$ and  $E_{\rm L}$-$A_{\rm T}$.\\
Here $\Xi_{1,2,3}$ depend on the opening angle $\beta$ and $y\equiv 2\pi D f c^{-1}$ ($D$: LISA-Taiji distance), and written by  trigonometric functions and the spherical Bessel functions $j_{\rm n}(y)$ \cite{Flanagan:1993ix,Seto:2008sr}.  For example, we have \cite{Seto:2008sr}
\beq
\Xi_3(\beta,y)=\sin\frac{\beta}2 \lkk \lmk   -j_1+\frac78 j_3\rmk+ \lmk   j_1+\frac38 j_3\rmk\cos\beta   \rkk .
\eeq
In Fig. 3, we present the overlap reduction functions in the low frequency regime. The period of the wavy profile is roughly given by 
$c/D\sim 3$\,mHz, reflecting the spherical Bessel functions.  {The functions decay rapidly $\psim f^{-1}$  at $f\gg c/D$ as the phase coherence is lost by the positional difference. } Below, we use simplified notations such as $\gamma_{IA_{\rm L}E_{\rm T}}=\gamma_{IAE}$.

\begin{figure}
 \includegraphics[width=8.5cm]{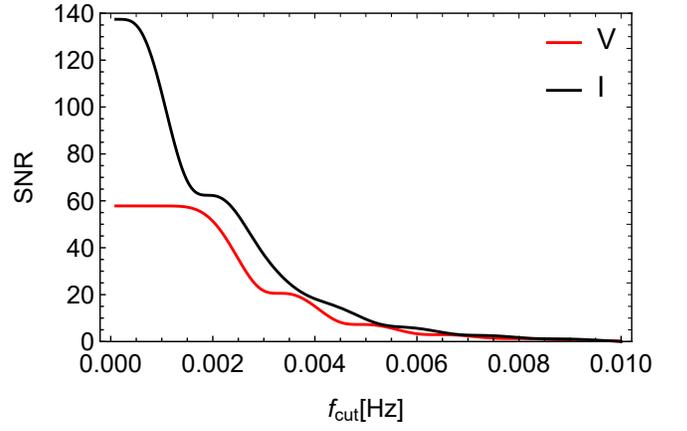}
 \caption{Signal-to-noise ratios for the $I$ and $V$ mode searches as functions of the minimum frequency $f_{\rm min}$ for signal integration (with  $f_{\rm max}=10$\,mHz and $T_{\rm obs}= 10$\,yr).  For signal models, we used $\Omega_{\rm GW}(f)=10^{-11}$ for the symmetric component and  $\Pi(f)\Omega_{\rm GW}(f)=10^{-11}$ for the asymmetric one. }
 \label{fig:4}
\end{figure}


\subsection{detection prospects}
Following the standard arguments on the correlation analysis \cite{Flanagan:1993ix,Allen:1997ad}, the signal-to-noise ratios $SNR_X$  for two modes  ($X=I,V$) are given by
\beq
SNR_X^2=\lmk \frac{16\pi}5  \rmk^2 T_{\rm obs}  \lkk 2\int_{f_{\rm min}}^{f_{\rm max}}  df \frac{\Gamma_X(f) X(f)^2}{f^6S_{\rm L}(f)S_{\rm T}(f)}   \rkk   \label{snr}
\eeq
with $\Gamma_I=\gamma_{IAA}^2+\gamma_{IEE}^2$ and $\Gamma_V=2\gamma_{VAE}^2$ due to the optimal parity decomposition.  
Here $T_{\rm obs}$ is the observational period, $(f_{\rm min},f_{\rm max})$ is the  frequency range for signal integration, and $S_{L,T}(f)$ are the instrumental noise spectra of the two detectors (shown in Fig. 1) without including confusion noise in the present  calculations.

For model characterization, we use the relations  $[I(f),V(f)]=\rho_c/(4\pi^2f^3) [\Omega_{\rm GW}(f),\Pi(f)\Omega_{\rm GW}(f)]$ with the critical density $\rho_c$.  In Fig. 4, we present $SNR_X$  ($X=I,V$) respectively  for  the fiducial background $\Omega_{\rm GW}=10^{-11}$ and $ \Pi \Omega_{\rm GW}=10^{-11}$ without $f$-dependence. 
We put $T_{\rm obs}=10$\,yr and   $f_{\rm max}=10$\,mHz, but changed $f_{\rm min}$. 
For the numerical values in Fig. 4, we have  
 simple scalings 
\[
\propto\lmk \frac{\Omega_{\rm GW}}{10^{-11}}\rmk  \lmk \frac{T_{\rm obs}}{10{\rm\, yr}}\rmk^{1/2},~~ \propto \lmk \frac{\Omega_{\rm GW}\Pi} {10^{-11}}\rmk  \lmk \frac{T_{\rm obs}}{10{\rm \,yr}}\rmk^{1/2}
\]
for $SNR_I$ and $SNR_V$.  
From  these scaling relations, the detection limits ($SNR_X\sim 5$) are estimated to be $\Omega_{\rm GW}\sim 10^{-12}$ and $\Pi \Omega_{\rm GW}=10^{-12}$ ( for $f_{\rm min}=2$\,mHz and $T_{\rm obs}=10$\,yr). 
 {In  the case of a flat spectrum, using the numerical results for $f_{\rm min}=2$\,mHz, the detection limit of the polarization degree ($|\Pi|<1$) is given by }
\beq
|\Pi|>0.08 \lmk \frac{\Omega_{GW}}{10^{-11}} \rmk^{-1} \lmk \frac{T_{\rm obs}}{10{\rm yr}}\rmk ^{-1/2} \lmk \frac{SNR}{5}   \rmk . 
\eeq

 Interestingly, we can observe   stair-like structures in Fig. 4,  reflecting the shapes of the overlap reduction functions. For  example, if we decrease $f_{\rm min}$  from 3\,mHz to 2\,mHz, $SNR_V$ and $SNR_I$ become 2.4 and 1.7 times larger.  We will also have a significant increase of $SNR_I$ by decreasing  $f_{\rm min}$ further below 2\,mHz, in contrast to $SNR_V$.  

{If we change  $f_{\rm max}$ from 10\,mHz to $30\,{\rm mHz}(>c/2\pi L)$, keeping $f_{\rm min}=$2\,mHz, the numerical values $SNR_V$ in  Eq. (11) change less than 0.012\% and 0.2\% respectively for $\Omega_{GW}|\Pi|\propto f^0$ and $f^1$ (with similar corrections for $SNR_I$). In Eq. (11), this weak dependence on $f_{\rm max}$ is  due to the factor $f^{-6}$ and the suppression of the overlap reduction functions, and  justifies our low frequency approximation for estimating $SNR_{I,V}$ (except for heavily blue-tilted spectra).  }

\section{discussion}

In this paper, we discussed correlation analysis for an isotropic gravitational wave background with heliocentric interferometers such as  
the LISA-Taiji network, paying special attention to the two parity modes  and the underlying geometrical  symmetries (see Fig. 2). 
Our analysis can be straightforwardly applied to a network composed by more than two heliocentric triangles detectors.

By correlating LISA and Taiji for ten years, our detection limit could reach $|\Pi|\Omega_{\rm GW}\sim  10^{-12}$ that is  four orders of magnitude better than the expected level with current generation ground-based detector network in the near future.  {This is enough sensitivity for examining some of theoretical predictions  including chromo-natural inflation discussed in \cite{Obata:2014loa}).}
Owing to the clear parity decomposition, we might uncover a parity violation signature in  a cosmological background, even if its energy density is dominated  by astrophysical confusion noise.

 As shown in Fig. 4, the estimated $SNR_{I,V}$ depend interestingly on the minimum frequency $f_{\rm min}$ for the signal integration. This frequency will be closely related to the processing status of the  Galactic binary subtraction, and the observational time $T_{\rm obs}$ is the key strategic parameter (see Fig. 1). The foreground subtraction would be a crucial aspect for the follow-on space projects such as DECIGO and BBO targeting weak primordial background around 0.1-1\,Hz \cite{Cutler:2005qq,Kawamura:2006up} (see also  \cite{Seto:2006dz} for potential orbital adjustment for the $V$ parameter search). The correlation analysis at mHz range would be useful also to examine the quality of the Galactic binary subtraction.

\begin{acknowledgments}
The author would like to thank H. Omiya for useful conversations. 
 This work is supported by JSPS Kakenhi Grant-in-Aid for Scientific Research
 (Nos. 17H06358 and 19K03870).
\end{acknowledgments}

\end{document}